\documentclass[aps,prl,twocolumn,groupedaddress,nofootinbib,notitlepage,showpacs,floatfix]{revtex4-1}

\usepackage{graphicx,graphics,epsfig,subfigure,times,bm,bbm,amssymb,amsmath,amsfonts,amsthm,mathrsfs,MnSymbol}
\usepackage[matrix,frame,arrow]{xypic}
\usepackage[pdfstartview=FitH]{hyperref}
\hypersetup{
}
\usepackage[pdftex]{color}
\usepackage{hypernat}
\usepackage{braket}
\usepackage{enumerate}
\usepackage[normalem]{ulem}

\newcommand{\bes} {\begin{subequations}}
\newcommand{\ees} {\end{subequations}}
\newcommand{\bea} {\begin{eqnarray}}
\newcommand{\eea} {\end{eqnarray}}

\newcommand{\eps}{\ensuremath{\varepsilon}} 

\newcommand{\beq}{\begin{equation}}

\newcommand{\eeq}{\end{equation}}

\newcommand{\ignore}[1]{}
\newcommand{\mc}[1]{\mathcal{#1}}

\def\b{\beta}

\def\s{\sigma}

\def\>{\rangle}
\def\<{\langle}

\newcommand{\ketbra}[1]{|{#1}\>\<#1|}

\def\s0{I}

\def\eps{\epsilon}

\newcommand{\ig}[1]{}

\def\dgr{\dagger}

\begin{document}
\title{High Fidelity Adiabatic Quantum Computation via Dynamical Decoupling}
\author{Gregory Quiroz}
\affiliation{Department of Physics and Center for Quantum Information Science \&
Technology, University of Southern California, Los Angeles, California
90089, USA}
\author{Daniel A. Lidar}
\affiliation{Departments of Electrical Engineering, Chemistry, and Physics, and Center
for Quantum Information Science \& Technology, University of Southern
California, Los Angeles, California 90089, USA}
\begin{abstract}
We introduce high-order dynamical decoupling strategies for open system adiabatic quantum computation. 
Our numerical results demonstrate that a judicious choice of high-order dynamical decoupling method, in conjunction with an encoding which allows computation to proceed alongside decoupling, can dramatically enhance the fidelity of adiabatic quantum computation in spite of decoherence.
\end{abstract}
\maketitle
\textit{Introduction.}---%
In adiabatic quantum computation (AQC) a problem is solved by evolving in the ground state manifold of an adiabatic Hamiltonian $H_{\textrm{ad}}(t)$, with $t\in[0,T]$ \cite{FarhiAQC:00,Farhi:01}. The ground state of the beginning Hamiltonian $H_B=H_{\textrm{ad}}(0)$ is assumed to be easy to prepare, while the final one, the ground state of the problem Hamiltonian $H_P=H_{\textrm{ad}}(T)$, represents the solution to the computational problem. 
AQC has been shown to be computationally equivalent to the standard circuit model of QC \cite{Aharonov:04,kempe:1070,Siu:2005:062314,Oliveira:05,PhysRevLett.99.070502}, and is being pursued experimentally using superconducting flux qubits \cite{Johnson:2011ys} and nuclear magnetic resonance \cite{Steffen:2003uq,Peng-AQC}. However, in spite of evidence of intrinsic robustness \cite{PhysRevA.65.012322,SarandyLidar:05,PhysRevA.71.032330,PhysRevA.75.062313,PhysRevA.79.022107,TAQC} and proposals to protect AQC against decoherence \cite{Jordan:2005:052322,LidarAQCDD:08}, AQC still lacks a complete theory of fault-tolerance, unlike the circuit model of QC \cite{Steane:2003:042322,Knill:2005:39,Terhal:2005:012336,Aharonov:2006:050504,Aliferis:2006:97,AliferisCross:07}.
In fact, even identifying an acceptable notion of fault-tolerant AQC (FTAQC) is an open problem. In the circuit model a fault-tolerant simulation allows one to generate the output of a given ideal circuit, to arbitrary accuracy, using the faulty components of another circuit \cite{Nielsen:book}. A similar definition for AQC would presumably involve the simulation of an ideal adiabatic evolution using a faulty one, but if ``faulty'' is simply taken to mean ``non-adiabatic", then one can just use the equivalence proof \cite{Aharonov:04,kempe:1070,Siu:2005:062314,Oliveira:05,PhysRevLett.99.070502} and circuit-model fault-tolerance \cite{Steane:2003:042322,Knill:2005:39,Terhal:2005:012336,Aharonov:2006:050504,Aliferis:2006:97,AliferisCross:07}
 to argue that the problem is already solved. However, this argument misses the point since to qualify as AQC, at least the \emph{computation} should remain adiabatic, i.e., the defining feature of AQC---the adiabatic preparation of the ground state of $H_P$---should be preserved. On the other hand, it seems too restrictive to require  the techniques used to address decoherence and noise to  be adiabatic as well.
We thus propose the following characterization of FTAQC:
`Given a closed-system AQC specified by $H_B$ and $H_P$, and $\epsilon>0$, a fault tolerant open-system simulation will use adiabatic evolution between two faulty, encoded Hamiltonians $\bar{H}_B$ and $\bar{H}_P$ derived from $H_B$ and $H_P$, so that the final system-only state of the simulation is efficiently decodable to a state that is $\epsilon$-close (in fidelity) to the ground state of $H_P$. In addition, the simulation may involve any other faulty non-adiabatic error-correction, suppression, or avoidance operations.'

Our characterization is meant to convey that the computation should be adiabatic, and apart from that the error correction can be anything. We make no attempt to rigorously quantify the relation between the ``ideal" and ``faulty" pairs $H_B,H_P$ and $\bar{H}_B,\bar{H}_P$, the types of allowed error-correction operations, or the nature of the decoding step. Instead, we demonstrate here that it is possible to approach ideal AQC in an open system, using ideas guided by our characterization of FTAQC. Specifically, we shall assume that the underlying \emph{computation} is indeed adiabatic but encoded into new Hamiltonians, and that the \emph{protection} is non-adiabatic. 

Ref.~\cite{LidarAQCDD:08} introduced an AQC protection method which fits this FTAQC approach. Protection is carried out by means of dynamical decoupling (DD) \cite{Viola:1998:2733, Zanardi:1999:77,Viola:1999:2417}, a non-adiabatic technique which involves the application of strong and frequent pulses to the system, designed to decouple it from the environment, or bath. To ensure the compatibility of DD with AQC, all qubits are encoded into a quantum error detecting code \cite{Rains:1999:266}, which allows DD pulses to be applied that commute with the adiabatic evolution, while at the same time acting to decouple the system from the bath \cite{comment-equiv}. Ref.~\cite{LidarAQCDD:08} relied on first order DD sequences, resulting in a tradeoff between fidelity and DD sequence bandwidth, and conjectured that high-order DD, in particular concatenated DD (CDD) \cite{Khodjasteh:2005:180501}, should alleviate this tradeoff. 

Here we demonstrate, using numerical simulations, that following the strategy of Ref.~\cite{LidarAQCDD:08}, but using high-order DD sequences, and in particular CDD, it is possible to dramatically enhance the fidelity of AQC in an open system setting. Our results support the idea that FTAQC, in the sense characterized above, is indeed attainable using appropriately chosen DD sequences.

\textit{Algorithms}.---%
We consider two well-known AQC algorithms, which correspond to first and second order quantum phase transitions, respectively, and thus to different challenges for AQC \cite{Schuetzhold:2006kx,PhysRevA.79.022107,PhysRevA.80.062326}.
The first is 
Grover's algorithm for the identification of a marked element in an unsorted list of $N$ elements, using the minimum number of oracle queries \cite{Grover:1997:325}. This can be done in ${O}(\sqrt{N})$ queries, which is a {quadratic} improvement over the best possible classical algorithm \cite{Bennett:1997fk}. Recast in the language of AQC \cite{GroverRC:02,RKHLZ:09}, Grover's algorithm is defined by the $n-$qubit Hamiltonian
$H^{\textrm{G}}_{\textrm{ad}}(t)=[1-f(t)](I-\ketbra{u})+f(t)(I-\ketbra{m})$,
where $\ket{u}$ denotes the uniform superposition over all $N=2^n$ computational basis states, $\ket{m}$ is the marked state, and $I$ is the identity operator. The minimum spectral gap $\Delta^{\textrm{G}}_{\min}={O}(1/\sqrt{N})$, and the total run time required to reach the ground state $\ket{\Phi^{\textrm{G}}_0(T)}=\ket{m}$ of $H^{\textrm{G}}_{\textrm{ad}}(T)$ is $T\sim O(\sqrt{N})$ provided the optimized interpolation function $f(t)=\frac{1}{2}-\frac{1}{2\sqrt{N-1}}\tan[(1-2t/T)\arccos(1/\sqrt{N})]$ is used \cite{GroverRC:02,RKHLZ:09}.

The second algorithm we consider is 2-SAT on a ring \cite{FarhiAQC:00}. Given a set of clauses $\{C_j\}^{n}_{j=1}$ associated with $n$ bits, such that each clause acts only on adjacent bits, the 2-SAT Hamiltonian acquires minimum energy when there are satisfying assignments for all clauses. In the case that the clauses define agreement between adjacent bits ($00$ or $11$, but not $01$ or $10$), the 2-SAT Hamiltonian is that of the transverse-field Ising model with periodic boundary conditions $\sigma^z_{n+1}\equiv\sigma^z_1$:
$H^{\textrm{2SAT}}_{\textrm{ad}}(t)=\left(1-\frac{t}{T}\right)\sum^{n}_{j=1}(I-\sigma^{x}_{j})+\frac{t}{T}\sum^{n}_{j=1}\frac{1}{2}(I-\sigma^{z}_{j}\sigma^{z}_{j+1})$.
Here $\sigma^{x}_j$ and $\sigma^{z}_j$ are the spin-$1/2$ Pauli matrices acting on qubit $j$. The symmetric ground state of $H^{\textrm{2SAT}}_{\textrm{ad}}(T)$, $\ket{\Phi^{\textrm{2SAT}}_0(T)}=(\ket{0\ldots 0}+\ket{1\ldots 1})/\sqrt{2}$, satisfies the clause agreement condition. The minimum gap occurs near $t=2T/3$ and $\Delta_{\textrm{min}}={O}(1/n)$; it can be found using the standard fermionization method \cite{FarhiAQC:00,Sachdev:1999:CambridgeUniversityPress}. The linear interpolation in 
$H^{\textrm{2SAT}}_{\textrm{ad}}(t)$ is not optimal; see Ref.~\cite{Rezakhani:2010fk} for the optimal path. The 2-SAT on a ring problem is not associated with a quantum speedup (there exists a classical linear-time algorithm for 2-SAT \cite{2SAT}), but is instructive nonetheless, and important in its own right as it corresponds to the preparation of the maximally entangled cat state $\ket{\Phi^{\textrm{2SAT}}_0(T)}$.

\textit{Error model.}---%
To keep our numerical simulations manageable, we considered interactions with a classical environment. Thus the ``faulty Hamiltonian" is
\begin{equation}
H_0(t)=H_{\textrm{ad}}(t)+H_{{\textrm{err}}}(t),
\label{eq:errModel}
\end{equation}
where the error Hamiltonian comprises interactions between the qubits and time-dependent classical fields:
\begin{equation}
H_{{\textrm{err}}}(t)=\sum_{\mu\in\{x,y,z\}}\sum_{j=1}^n\eps^{\mu}_j(t)\,\sigma^{\mu}_j .
\label{eq:He}
\end{equation}
Each field $\eps^{\mu}_j(t)$ is a zero-mean random Gaussian process 
with 
spectral density $S_{jk}^{\mu\nu}(\omega) = \delta_{jk}\delta_{\mu\nu}S(\omega)$ \cite{Ricker:book}, where 
\begin{equation} 
S(\omega)=\frac{1}{\sqrt{2\pi}}\int^{\infty}_{-\infty}\left\langle \eps^{\mu}_j(t)\eps^{\mu}_j(t+\tau)\right\rangle e^{i\omega\tau}d\tau = {\frac{\b}{\sqrt{2\pi}}}e^{-\frac{1}{2}\left(\frac{\omega}{\b}\right)^2},
\label{eq:crossCor}
\end{equation}
where $\left\langle\cdot\right\rangle$ denotes a Gaussian ensemble average. Note that $\beta$ plays the role of the spectral cutoff ($1/\b$ is the bath correlation time).
While $H_{{\textrm{err}}}(t)$ is not as general as a quantum environment, it still represents an interesting and relevant error model, e.g., due to charge noise in superconducting qubits \cite{Galperin:2006fk,Cywinski:2008:174509}: its terms combine with $H_{\textrm{ad}}(t)$ to produce faulty operations by inducing random time-dependent fluctuations in the spectrum of $H_{\textrm{ad}}(t)$. In particular, these fluctuations also act to modify the size and location of the minimum energy gap $\Delta_{\textrm{min}}$ of $H_{\textrm{ad}}$.

\textit{Dynamical decoupling.}---The problem we attempt to solve using DD is to perform high fidelity AQC in spite of the presence of $H_{\textrm{err}}$. The decoupling pulses are introduced through an additional time-dependent control Hamiltonian $H_C(t)$, which generates a unitary pulse propagator $U_C(t)$. We consider zero-width pulses separated by finite intervals. The inclusion of pulse-width errors is left for a future study focusing on a more complete picture of fault tolerance; CDD is known to be relatively robust against such errors \cite{Khodjasteh:2005:180501,Ng:2011:012305,Suter-CDD}.  The Hamiltonian $H(t)=H_C(t)+H_0(t)$ generates the complete dynamics of the system in the presence of DD, represented by the unitary evolution operator $U(t,0)$. To suppress $H_{{\textrm{err}}}(t)$, while preserving $H_{\textrm{ad}}(t)$, we require that each term in $H_{{\textrm{err}}}(t)$ anticommute with some pulse operator comprising $H_C(t)$, while $[H_C(t),H_{\textrm{ad}}(t')]=0$ $\forall t,t'$. Upon satisfying these conditions the time evolution operator in the interaction (``toggling") frame with respect to $H_C(t)$ is
\bea
\tilde{U}(T,0) &:=& U_C^\dgr(T)U(T,0) \\
&=&\mathcal{T}e^{-iT\int^{1}_{0}H_\textrm{ad}(s)ds}+{O}[(\|\tilde{H}^{'}_{\textrm{err}}\|T)^{\alpha+1}],\notag
\eea
where $\mathcal{T}$ denotes time ordering, and $\tilde{H}^{'}_{\textrm{err}}$ is an effective error Hamiltonian, which can be computed using the Magnus or Dyson series \cite{Blanes:2009:151}. DD becomes effective provided the ``noise strength'' $\|\tilde{H}^{'}_{\textrm{err}}\|T<1$ \cite{Ng:2011:012305}. The larger the ``decoupling order'' $\alpha$, the closer to ideal is the adiabatic evolution. Previously only the case $\alpha=1$ was analyzed \cite{LidarAQCDD:08}; we shall show that protection via high-order DD sequences, in particular CDD,
alleviates this tradeoff.

\textit{Stabilizer decoupling.}---As in Ref.~\cite{LidarAQCDD:08}, to satisfy the non-interference condition $[H_C(t),H_{\textrm{ad}}(t')]=0$ we make use of  the $[[n,n-2,2]]$ stabilizer code $\mathcal{C}$, encoding $n-2$ logical qubits ($n$ even) into $n$ physical qubits \cite{Rains:1999:266,Viola:2000:3520}. The stabilizer of $\mathcal{C}$ is $\mathcal{S}=\{I,X,Y,Z\}$, where $X(Y,Z)=\bigotimes^{n}_{j=1}\sigma^{x(y,z)}_j$. The encoded single-qubit operators are $\bar{\sigma}^x_j=\sigma^{x}_{1}\sigma^{x}_{j+1}$ and $\bar{\sigma}^z_j=\sigma^{z}_{j+1}\sigma^{z}_{n}$, where $j=1,2,\ldots,n-2$. AQC over $\mathcal{C}$ is implemented by replacing each Pauli matrix in $H_{\textrm{ad}}(t)$ by its encoded version, yielding an encoded adiabatic Hamiltonian $\bar{H}_{\textrm{ad}}(t)$ which is fully $2$-local.

\textit{CDD for AQC.}---Consider a unitary decoupling group $\mathcal{G} = \{g_k\}_{k=1}^G$, chosen to implement first-order decoupling ($\alpha=1$). 
DD effectively averages out $H_{\textrm{err}}$ by symmetrizing it over $\mathcal{G}$ \cite{Zanardi:1999:77,Viola:1999:2417}. We choose $\mathcal{G}=\mathcal{S}$ and hence $G=4$ for the $[[n,n-2,2]]$ code. 
CDD achieves higher order decoupling by concatenating this symmetrization, so that each additional level averages out the remaining leading order. So far CDD has been defined only for piecewise constant Hamiltonians \cite{Khodjasteh:2005:180501,Ng:2011:012305,KLCDD:07}. Adapting CDD to incorporate the time-dependence of $H(t)$, we have at the $l+1$th level of concatenation
\beq
\label{eq:CDD}
U_{\rm CDD}^{(l+1)}(T,0) = \prod_{k_1=1}^G g_{k_1} P^{(l)}_{k_1} g_{k_1}^\dgr,\quad l\geq 0 .
\eeq
Starting from $m=2$, $P^{(l)}_{k_1}$ is calculated recursively from
\beq
P^{(l)}_{k_1,\ldots,k_{m-1}} := \prod_{k_m=1}^G g_{k_m} P^{(l-1)}_{k_1,\ldots,k_m} g_{k_m}^\dgr ,\quad l\geq 1
\eeq
with DD-free evolution segments
\begin{align}
\begin{split}
P^{(0)}_{k_1,\ldots,k_l} &:= U_0\left(t_{l-1}+T\frac{k_l}{G^l},t_{l-1}+T\frac{k_l-1}{G^l}\right)  \\
t_l &:= T\sum_{j=1}^{l}\frac{k_j}{G^j},\quad t_0=0,\quad l\geq 1,
\end{split}
\end{align}
where the DD-free evolution operator is $U_0(t_a,t_b)=\mathcal{T}\exp\left(-i\int^{t_a}_{t_{b}}dt H_0(t)\right)$. These definitions ensure that at concatenation level $l$ the symmetrization procedure is applied to DD-free evolution segments of duration $\tau_l = T/G^l$, using a total of $G^l$ pulses. For piecewise constant Hamiltonians it has been shown that CDD achieves $\alpha=l$th order decoupling \cite{Ng:2011:012305,KLCDD:07,Witzel:2007fk}. Experimental studies support these conclusions \cite{Peng-CDD,Suter-CDD}.

\begin{figure}[t]
\includegraphics[width=\columnwidth]{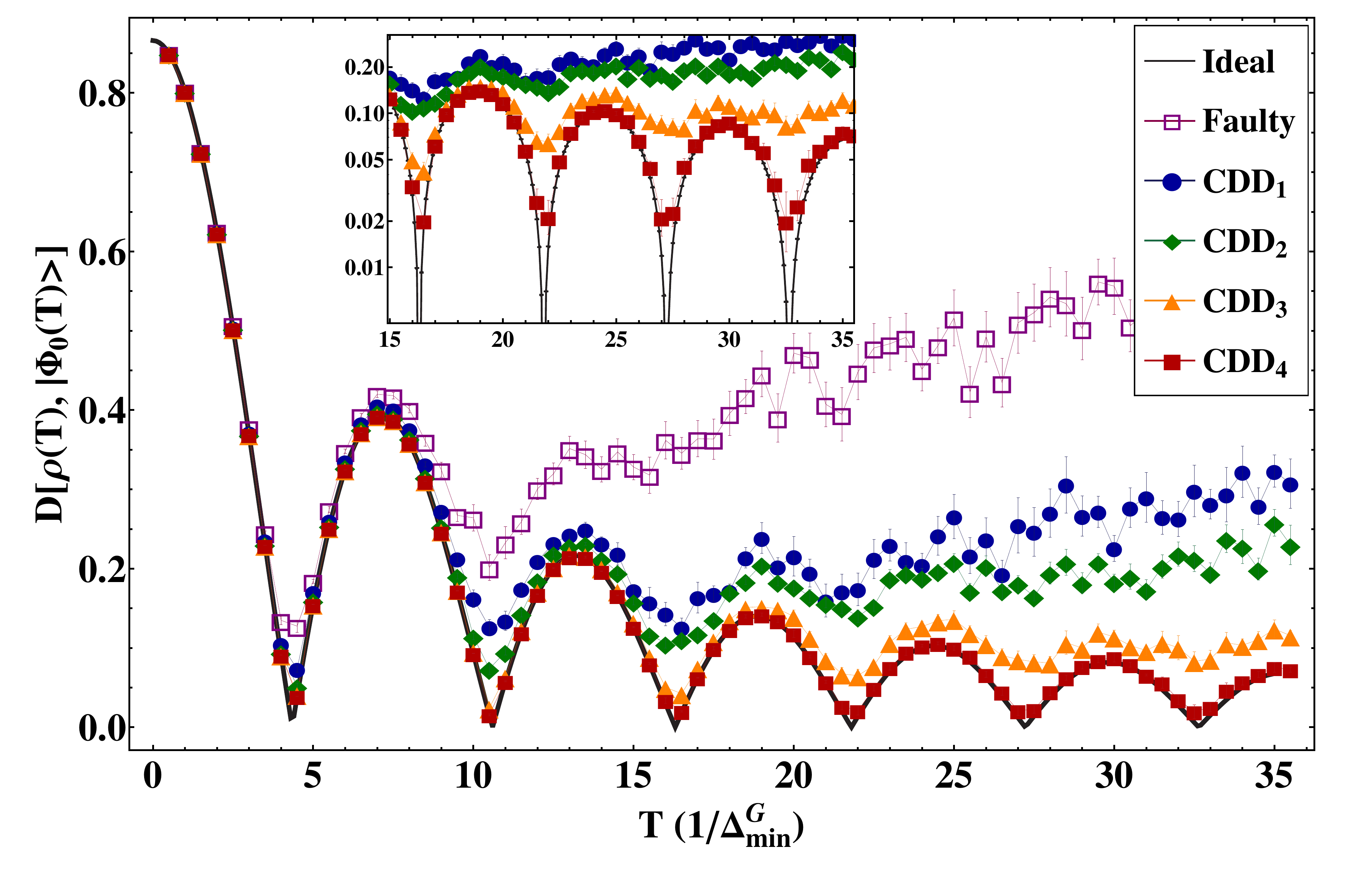}
\caption{(color online) Trace-norm distance between the CDD-protected final state $\rho(T)$ and the desired ground state $\ket{\Phi_0(T)}$, as a function of total run time $T$ for Grover's algorithm, in units of the inverse minimum gap. Cutoff frequency: $\beta=\Delta/5$. The ideal (solid black) and faulty (empty squares) evolutions are included for reference. Insert shows a close-up for large $T$, using a log scale for the vertical axis. Performance improves monotonically with concatenation level $l$, with the corresponding sequence denoted CDD$_l$. Error bars are due to averaging over 30 random realizations of $H_{\textrm{err}}(t)$.}
\label{fig:GroverCDD}
\end{figure}

\begin{figure}[t]
\includegraphics[width=\columnwidth]{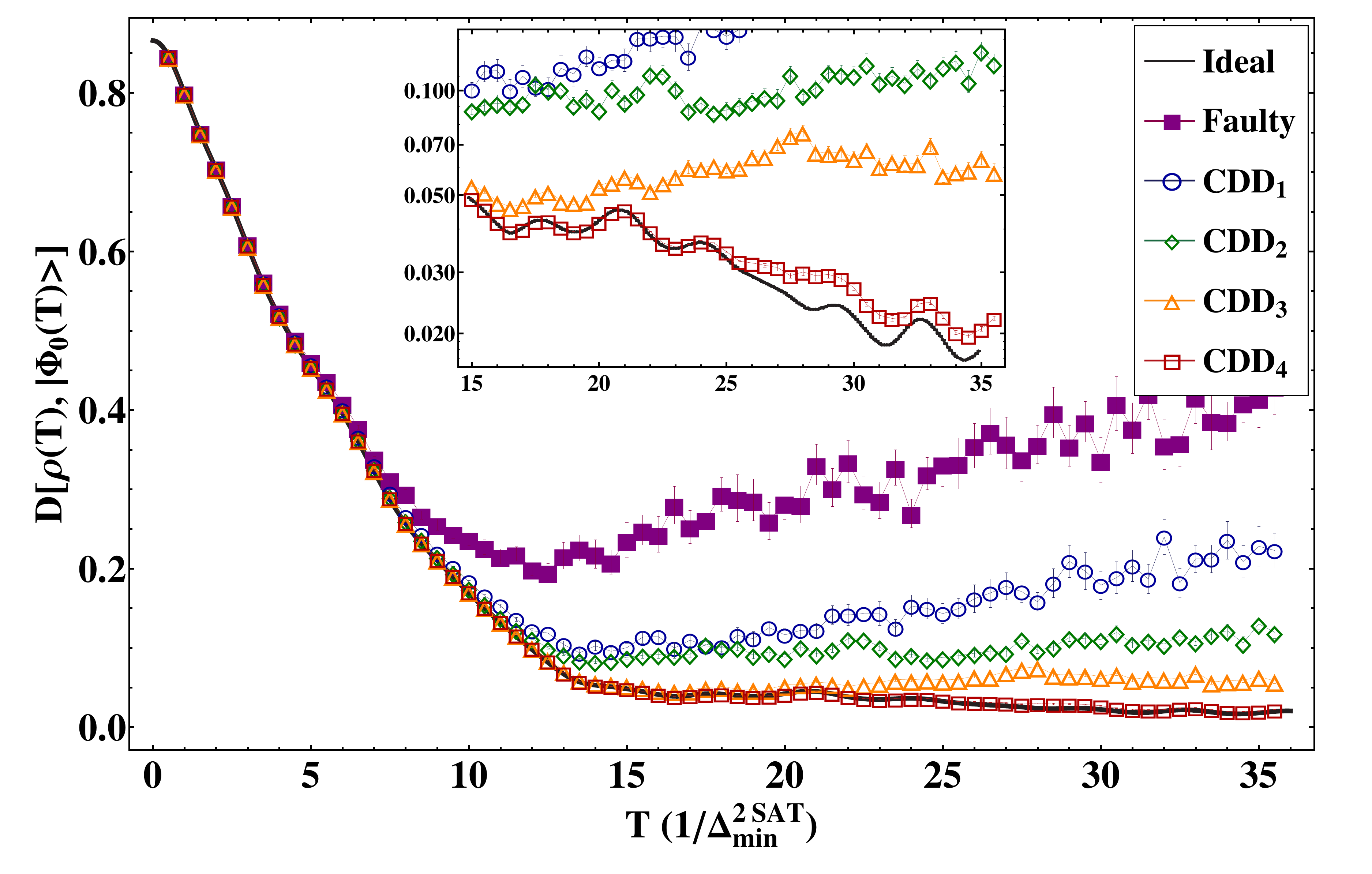}
\caption{(color online) As in Fig.~\ref{fig:GroverCDD}, 
for the 2-SAT on a ring problem. The curves for ideal evolution and CDD$_4$ overlap to within our numerical accuracy up to $T=25/\Delta_{\min}^{\rm 2SAT}$. Note the minimum at $T\approx 12 \Delta_{\min}^{\rm 2SAT}$ for the faulty evolution, suggesting the existence of an optimal open system evolution time. This optimal time increases with concatenation level, until it disappears in the ideal case and for CDD$_4$. CDD boosts the deviation by a factor of $10$ from $D\approx 0.2$ (at $T\approx 12.5/\Delta^{{\rm 2SAT}}_{\textrm{min}}$) in the faulty case to $D\approx 0.02$ (at $T\approx34.5/\Delta^{{\rm 2SAT}}_{\textrm{min}}$)
for CDD$_4$.}
\label{fig:2SATCDD}
\end{figure}

\textit{QDD for AQC.}---Another way to achieve high order decoupling in a multi-qubit setting is QDD \cite{West:2010:130501}, which is obtained by nesting two Uhrig DD \cite{Uhrig:2007:100504} (UDD)
 sequences
\bes
\begin{align}
\label{eq:QDD}
U^{(M_1,M_2)}_{\textrm{QDD}}(T,0) &= U^{(M_2)}_{\textrm{UDD},\Omega_2}\circ U^{(M_1)}_{\textrm{UDD},\Omega_1}\circ U_0(T,0) , \\ 
U^{(M)}_{\textrm{UDD},\Omega}\circ U_0(T,0) &= \Omega^{M+1}\prod^{M+1}_{k=1}\Omega U_0(\delta^{(M)}_k,\delta^{(M)}_{k-1}),
\end{align}
\ees
where $\delta^{(M)}_k=T\sin^{2}[k\pi/(2M+2)]$ and $\Omega_1\neq \Omega_2 \in\{X,Z\}$ are the generators of $\mc{S}$. The number of pulses in each sequence, or sequence order, is $\{M_1,M_2\}$, and dictates the decoupling order $\alpha$, requiring $(M_1+1)(M_2+1)$ total pulses for $\alpha \geq \min(M_1,M_2)$ \cite{Quiroz:2011kx,Kuo:1106.2151,Jiang:2011fk}. 
In principle 
QDD uses exponentially fewer pulses than CDD to attain the same decoupling order, but as we shall see in fact
CDD performs better in the context of our simulations of DD-protected AQC.

\begin{figure}[tp]
\includegraphics[width=\columnwidth]{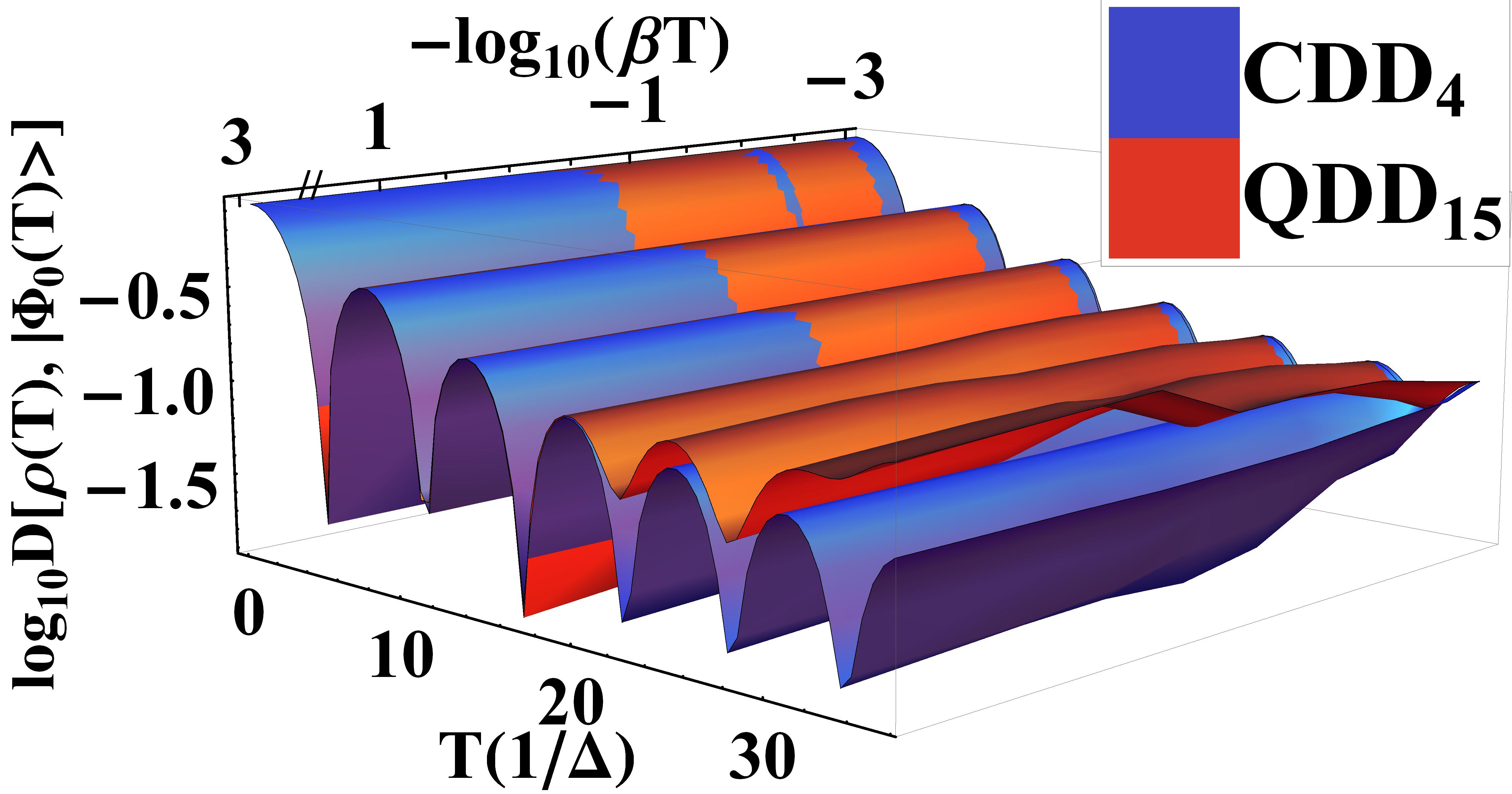}
\caption{(color online) CDD$_4$ \textit{vs} QDD$_{15}$ for the Grover problem  as a function of the normalized bath correlation time $1/(\b T)$ and normalized total run time $T/\Delta^{\textrm{G}}_{{\textrm{min}}}$. Increasing the bath correlation time $1/\b$ at fixed $T$ generally results in improved performance for CDD$_4$. whose performance is significantly better than QDD$_{15}$ in in the large $T$ regime. All results were averaged over $30$ realizations of $H_{\textrm{err}}(t)$.}
\label{fig:DistvsCorrTimevsTotalTGrover}
\end{figure}

\textit{Results.}---%
For each algorithm (Grover, 2SAT) we used $n=4$ physical qubits to encode two logical qubits in the code $\mathcal{C}$, and studied both CDD- and QDD-protected AQC.
In our simulations the pulse intervals decrease as the total number of pulses increases with the concatenation or QDD sequence order, for each given value of the total time $T$. Starting for both algorithms from the uniform superposition state as the initial encoded state $\ket{\psi(0)}$, we computed $\ket{\psi(T)} = U_{{\textrm{xDD}}}(T,0)\ket{\psi(0)}$, where x$=$C or Q [Eqs.~\eqref{eq:CDD} and \eqref{eq:QDD}], for a given Gaussian noise realization. Computing the ensemble average $\rho(T)=\<\ketbra{\psi(T)}\>$, we assessed performance using the trace-norm distance $D[\rho(T),\ket{\Phi_0(T)}]$ \cite{comment-TND}, to quantify the difference between the encoded, xDD-protected state $\rho(T)$
and the desired encoded final state $\ket{\Phi_0(T)}$, i.e., the ground state of $\bar{H}_{\textrm{ad}}(T)$. %
In the ideal case of noise- and DD-free evolution $\ket{\psi_{\textrm{ad}}(t)} = \mathcal{T}\exp[-\int_0^t\bar{H}_{\textrm{ad}}(t')dt']$, the adiabatic theorem \cite{Teufel:book}
guarantees $D[\ket{\psi_{\textrm{ad}}(T)},\ket{\Phi_0(T)}]\ll 1$ provided $T \gg \max_{s\in\{0,1\}}\|\frac{d}{ds}{\bar{H}}_{\textrm{ad}}\|^{b-1}/\Delta_{\min}^b$, $s=t/T$, $b\in\{1,2,3\}$ \cite{Jansen:07,lidar:102106,Wiebe:12}.
In our simulations a finite range of $T$'s was used, and as can be seen from Figs.~\ref{fig:GroverCDD}-\ref{fig:DistvsCorrTimevsTotalTGrover}, $D[\rho(T),\ket{\Phi_0(T)}]$ does indeed tend to zero for the ideal case as $T$ is increased, though not monotonically in the Grover case  \cite{comment-oscillations}. The main effect of  $H_{\textrm{err}}(t)$ is to cause $D[\rho(T),\ket{\Phi_0(T)}]$ to diverge away from zero as $T$ is increased, so that there is an optimal evolution time; this effect of open system AQC, due to a competition between the benefit of increasing $T$ for adiabaticity and the simultaneous increasing damage due to system-bath coupling, has been previously pointed out experimentally \cite{Steffen:2003uq} and theoretically \cite{Sarandy:2005fk,SarandyLidar:05}. The main role of DD protection, then, is to keep the fidelity of the AQC process as close as possible to the ideal and, in particular, to prevent $\min_{T\in[0,T_{\max}]} D[\rho(T),\ket{\Phi_0(T)}]$ from growing larger than some tolerance $\epsilon>0$ away from $\min_{T\in[0,T_{\max}]} D[\ket{\psi_{\textrm{ad}}(T)},\ket{\Phi_0(T)}]$. This is the sense in which DD-protected AQC approaches the ideal of FTAQC described in the introduction.

CDD results 
for Grover's problem are shown in Fig.~\ref{fig:GroverCDD}, for increasing concatenation levels, at $\beta = \Delta/5$.
The faulty Grover evolution [generated by $\bar{H}^{\textrm{G}}_{\textrm{ad}}(t)+H_{\textrm{err}}(t)$] reaches a minimum deviation $D\approx 0.13$ at  {$T\approx 4.5/\Delta^{\textrm{G}}_{\textrm{min}}$} and then diverges from the ideal evolution [generated by $\bar{H}^{\textrm{G}}_{\textrm{ad}}(t)$]. In contrast, CDD-protected evolution becomes remarkably close to the ideal evolution as the level of concatenation increases. Nearly ideal evolution is maintained essentially over the entire range of $T$ values we simulated. 

Figure~\ref{fig:2SATCDD} 
shows our CDD results 
for the 2-SAT problem. The results are even better than those for the Grover problem. CDD-protected evolution is essentially indistinguishable from the ideal at high enough concatenation level. 
The improved performance in the 2-SAT case can be attributed to the larger gap: $\Delta_{\min}$ scales inversely with the number of qubits in 2-SAT, instead of with the 
Hilbert space dimension as in the Grover case, rendering the 2-SAT evolution less sensitive to bath-induced noise effects. This is consistent with earlier observations that algorithms associated with second order quantum phase transitions are more amenable to AQC than those for first order transitions \cite{Schuetzhold:2006kx,PhysRevA.79.022107,PhysRevA.80.062326}.

Along with CDD, we analyzed QDD-protected AQC for $M_1=M_2=M\in\{1,3,6,7,14,15\}$, where the odd sequence orders correspond to the first four levels of concatenation in CDD$_l$, $l\in\{1,2,3,4\}$, respectively, having the same number of pulses [$(M+1)^2$ and $4^l$]. CDD and QDD are compared in our simulations at equal $T$ values. QDD$_M$ performance improves monotonically for sufficiently large $\b$ (not shown), but we find that QDD$_{15}$ 
performance is consistently inferior to that of CDD$_{4}$ in the adiabatic regime of large $T\Delta$, as can be seen in Fig.~\ref{fig:DistvsCorrTimevsTotalTGrover}. This is surprising in light of the aforementioned fact that in their non-AQC roles as quantum memory DD sequences, the respective decoupling orders of CDD$_l$ and QDD$_{M}$ are $l$ and $\geq M$. We have confirmed that the results for the 2SAT problem are qualitatively similar, again favoring CDD.

\begin{figure}[t]
\includegraphics[width=\columnwidth]{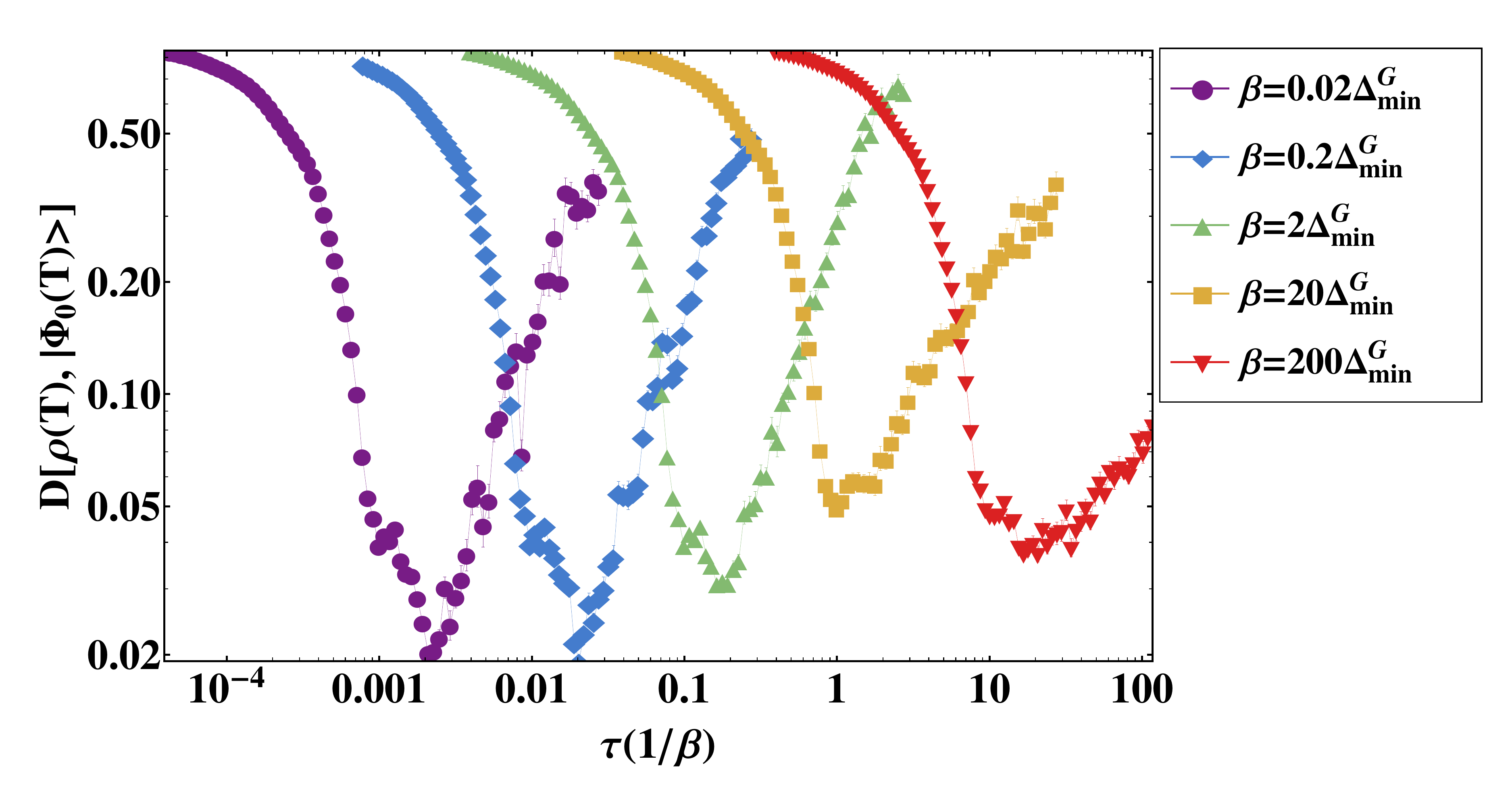}
\caption{(color online) Performance of CDD$_4$-protected evolution for the 2SAT problem, as a function of pulse interval $\tau$, for different values of the frequency cutoff $\b$ ($\tau$ in units of $1/\b$ to separate the curves). Total time $T= 4^4\tau$. The minimum in each fixed $\b$ curve corresponds to an optimal pulse interval $\tau_{\textrm{opt}}(\b)$, and corresponding $T_{\textrm{opt}}(\b)$. Peak performance $D[\rho(T_{\textrm{opt}}(\b)),\ket{\Phi_0(T_{\textrm{opt}}(\b))}]$ improves as $\b$ is decreased, except at $\beta=\Delta^{2SAT}_\textrm{min}/0.005$ where self-averaging effects result from rapid fluctuations in $\eps^{\mu}_j(t)$ (``motional narrowing").  Results are averaged over $30$ realizations of $H_{\textrm{err}}(t)$.}
\label{fig:2SATCDD4-DvsCorrT}
\end{figure}

In Figs.~\ref{fig:GroverCDD} and \ref{fig:2SATCDD} the cutoff frequency $\b$ was fixed. Performance dependence on $\b$ is shown in Fig.~\ref{fig:DistvsCorrTimevsTotalTGrover}, and is seen to be mild for CDD$_{4}$, at each fixed value of $T$. The dependence on $\b$ is further elucidated in Fig.~\ref{fig:2SATCDD4-DvsCorrT}, which shows that performance generally improves as $\b$ shrinks, as expected.
In contrast, QDD$_{15}$ results show almost no dependence on $\b$, and the $\min_{\tau}D[\rho(T(\tau)),\ket{\Phi_0(T(\tau))}]$ values are approximately twice as large as those obtained for CDD$_4$ (not shown). 
 In essence, the superior performance of CDD can be explained by recognizing that QDD is designed to suppress the system-bath coupling at the \emph{end} of the pulse sequence, while CDD performs this suppression recursively all throughout the evolution. This is a better fit for AQC, with its time-dependent system Hamiltonian. 
 
\textit{Conclusion and future work.}---%
We have introduced a high-order DD-based strategy for protected open-system AQC, and demonstrated using numerical simulations that it is capable of achieving high fidelities for a model of classical stochastic noise. At high enough concatenation level, the CDD-based protection strategy achieves fidelities which are essentially indistinguishable from closed-system adiabatic evolution, for two algorithms associated with first and second order quantum phase transitions.
Future work should elucidate the question of scaling of CDD resources with problem size, consider finite-width DD pulses, and generalize the results to a quantum bath. 

\textit{Acknowledgments}.---We are grateful to Dr. G. Paz-Silva for important discussions. This research was supported by the Department of Defense, the Lockheed Martin Corporation under the URI program, and by
NSF grants No. CCF-0726439, CHE-924318, CHM-1037992, PHY-969969, and PHY-803304. The views and conclusions contained in this document are those of the authors and should not be interpreted as representing the official policies, either expressly or implied, of the U.S. Government.


\bibliographystyle{apsrev4-1}

\begin{thebibliography}{64}
\bibitem [{com(){\natexlab{b}}}]{comment-TND}%
  \BibitemOpen
  \href@noop {} {}  \bibinfo {note} {\uppercase{T}he
  trace-norm distance $D[\rho_1,\rho_2]:=\frac{1}{2}\|\rho_1-\rho_2\|_1$, where
  $\|A\|_1:=\text{Tr}\sqrt{A^{\dagger}A}$, is a standard distance measure
  between quantum states \cite{Nielsen:book}. When $\rho_i$ is a pure state
  $\rho_i=\ketbra{\psi_i}$ we replace the corresponding argument of $D$ by
  $\ket{\psi_i}$.}\BibitemShut {Stop}%
\bibitem [{com(){\natexlab{c}}}]{comment-oscillations}%
  \BibitemOpen
  \href@noop {} {}  \bibinfo {note} {{Ideal case oscillations
  are consistent with the adiabatic theorem, which predicts $D\rightarrow 0$
  only in the limit $T \rightarrow \infty$. For an analytical expression
  predicting such oscillations in the case of Grover's algorithm see
  Ref.~\cite{RPL:10}, Eq.~(174).}}\BibitemShut {Stop}%
\end{thebibliography}

%

\end{document}